\documentclass[11pt, a4paper]{article}
\usepackage{latexsym,amsmath,amssymb,amsbsy,amsthm}

\newtheorem{theorem}{Theorem}[section]

\renewcommand\thesection{\Roman{section}.\!\!\!\!\!}
\makeatletter\renewcommand{\@biblabel}[1]{$^{#1}$\!\!\!}
\renewcommand{\@cite}[2]{{#1\if@tempswa, #2\fi}}
\renewcommand{\@makefnmark}{\hbox{\mathsurround=0pt $^{\textup a)}$}}

\newcommand{\refn}[1]{(\ref{#1})}
\newcommand{\hil}{\mathcal H}
\newcommand{\dom}{\mathcal D}
\newcommand{\kr}{\mathcal K}
\newcommand{\tp}{\tilde P}
\newcommand{\tq}{\tilde Q}
\newcommand{\tps}{\tilde\psi}
\newcommand{\citer}[1]{Ref.~\cite{#1}}
\newcommand{\citerm}[2]{Refs.~\cite{#1} and \cite{#2}}
\newcommand{\citev}[1]{$^{\scriptsize\cite{#1}}$}

\makeatother

\begin{document}
{\Large\bf Extension of von Neumann's uniqueness  theorem to the theories with non-physical particles }

\medskip
%\hspace*{1in}\parbox{114mm}{The Weyl\footnotemark\ form analogue in a Krein space with the anti-Fock representation of the Heisenberg algebra is shown to be unitarily equivalent to the Schr\"odinger representation.}\footnotetext{aa} $\footnote{Author to whom correspondence should be addressed.  Electronic mail:  kv.antipin@physics.msu.ru.}
\begin{quotation}\noindent
K.~V.~ Antipin,$^{1}$ M.~N.~Mnatsakanova,$^{2}$ and
Yu.~S.~Vernov$^{3}$\\
$^1$\small{\em Department of Physics, Moscow State University, Moscow 119991, Russia.} \\
$^2$\small{\em Skobeltsyn Institute of Nuclear Physics, Moscow State University, Moscow 119992, Russia.} \\
$^3$\small{\em Institute for Nuclear Research, Russian Academy of Sciences, Moscow 117312, Russia.}

\bigskip\bigskip\bigskip\smallskip\smallskip\medskip
\noindent Von Neumann's uniqueness theorem is extended to a special class of canonical commutation relations, namely the anti-Fock representations, which are realized  on a  Krein space.
\end{quotation}
\vspace{-0.5\baselineskip}

\section{Introduction}
\mbox{}\vspace{-\baselineskip}

The canonical commutation relations (CCR) form the basis of any quantum theory.
In the quantum mechanics of a single particle with one degree of freedom the canonical momentum and position variables
can be represented by self-adjoint operators $P$ and $Q$ on a Hilbert space $\hil$ satisfying
\begin{equation}\label{CCR}
[P,Q]=-iI.
\end{equation}
%\smallskip\medskip\bigskip\bigskip
%\vspace*{-3pt}
%\hrule width .4\columnwidth
%\vspace*{2.6pt}
%\noindent\hbox to 1.8em{\footnotesize $^{\textup a)}$ Author to whom correspondence should be addressed.  Electronic mail:  kv.antipin@physics.msu.ru.}\\\newpage
Note that the passage from one to a finite number of degrees of freedom, when canonical commutation relations are defined by
\begin{equation}\label{CCRm}
[P_i,Q_j]=-i\delta_{ij}I,\quad[Q_i,Q_j]=0,\quad[P_i,P_j]=0,\quad 1\leqslant i,j\leqslant n,
\end{equation}
does not involve any essential modification, and all the basic results for the CCR admit straightforward generalization\citev{Put}.

We say that a pair $(P, Q)$ of self-adjoint operators form a
Heisenberg representation of the CCR on a Hilbert space $\hil$ if
there exists a dense linear manifold $\dom$ in $\hil$ with
$\dom\subseteq\dom (P)\cap\dom (Q)$ such that relation~\refn{CCR}
holds. In~\citer{Put} it is shown that the operators $P$ and $Q$
are unbounded. Therefore, the corresponding domain questions arise
when we deal directly with $P$ and $Q$. The Weyl form of the CCR
representations is more convenient in this respect. It is given by
\begin{equation}\label{Weyl}
e^{itP}e^{isQ}=e^{ist}e^{isQ}e^{itP},\quad s,t\in\mathbb R.
\end{equation}
Indeed, according to Stone,\citev{Stone} the operators $e^{itP}$ and $e^{isQ}$ are bounded  and hence defined on entire space under consideration.

However, relation~\refn{Weyl} holds only for a certain class of unitarily equivalent representations that includes the Schr\"odinger one~(\citerm{Foi}{Vernreg}). The Schr\"odinger operators $p$, $q$ on $L^2(-\infty,\infty)$ are defined by
\begin{equation}\label{Sch}
(pf)(x)=-i\frac{\partial}{\partial x}f(x),\quad (qf)(x)=xf(x).
\end{equation}
Representations unitarily equivalent to the Schr\"odinger one are referred to as regular representations~(\citerm{Foi}{Vernreg}).

Note that there exist Heisenberg representations of the canonical commutation relation, even for one degree of
freedom, which are not unitarily equivalent to the Schr\"odinger representation, that~is, are irregular. For example, the operators~\refn{Sch}, which are defined on the space $L^2(a,b),\,\,a,b\in\mathbb R$, form an irregular representation of the CCR algebra.

Consequently, if we want to describe regular representations, we need to impose some
conditions that are more stringent than those required in the definition of a
Heisenberg representation; one way to achieve this is to impose conditions
on the domains and properties of the self-adjoint operators $P$ and $Q$. Several
results have been reported in this direction; among them we can mention that of Rellich\citev{Rell} and Dixmier\citev{Dix}.
\renewcommand{\thesection}{\arabic{section}}
\begin{theorem}
In order that a pair $(P, Q)$ of closed symmetric operators, acting
on a  Hilbert space $\hil$, form a regular Heisenberg representation of the CCR it is necessary and sufficient that there exists a
linear manifold $\dom$, contained in $\dom (P)\cap\dom (Q)$ and dense in $\hil$, such that
\begin{enumerate}
\item $\dom$ is stable with respect to $P$ and $Q$,
\item the restriction of $(P^2+Q^2)$ to $\dom$ is essentially self-adjoint,
\item $PQ-QP=-iI$ on $\dom$.
\end{enumerate}
\end{theorem}

The other way is to deal with the Weyl commutation relations~\refn{Weyl}. In this connection the following result is due to von Neumann\citev{Neumann}.
%It is known that von Neumann's theorem\citev{Neumann} guarantees the unitary equivalence of any two irreducible canonical systems with equal finite numbers of degrees of freedom.
\begin{theorem}[Von Neumann's theorem]
Let $\hil$ be a separable Hilbert space, $\{U(t)\mid t\in \mathbb R\}$ and $\{V(s)\mid s\in \mathbb R\}$ be two weakly continuous, one-parameter groups of unitary operators acting on $\hil$, such that
$$
U(t)V(s)=e^{its}V(s)U(t)\quad\forall t,s\in\mathbb R;
$$
then $\{U(t),\,V(s)\mid t,s\in \mathbb R\}$ is unitarily equivalent to a direct sum of Schr\"odinger representations.
\end{theorem}
\renewcommand\thesection{\Roman{section}.\!\!\!\!\!}

Note that relation~\refn{CCR} can be rewritten as follows
\begin{equation}\label{aa}
[a,a^*]=I,
\end{equation}
where
\begin{equation}\label{aa*}
a=\frac1{\sqrt2}(Q+iP),\quad a^*=\frac1{\sqrt2}(Q-iP).
\end{equation}
It is clear that the vacuum vector $\psi_0=C\mathrm{exp}(-x^2/2)$ satisfying
\begin{equation}\label{vac}
a\psi_0=0
\end{equation}
exists in the Schr\"odinger representation.

The representations for which relation~\refn{vac} holds are known as the Fock representations. It can be shown that all the Fock representations are unitarily equivalent.
In particular, every Fock representation is unitarily equivalent to the Schr\"odinger one and, therefore, is regular.

The CCR algebra representations in the Weyl form have mainly been investigated in Hilbert spaces\citev{BR}. However, in gauge quantum field theories, where, in a covariant gauge, the non-physical particles appear, it is necessary to consider spaces with an indefinite metric~(indefinite inner product)\citev{KO,MS,St}. Some analogues of the Weyl relations have recently been derived for an important class of indefinite metric spaces, namely the class of Krein spaces\citev{jmath}. It is this class of spaces that our paper will focus on. In~\citer{nonph} it is shown that, besides the Fock representation, two other types appear in Krein spaces, one with negative (the anti-Fock case), the other with two-sided discrete spectrum of the number operator $N=a^*a$. The aim of the present paper is to prove the applicability of von Neumann's theorem to the derived Weyl form analogue of the anti-Fock representation on a Krein space.

\section{Krein space}
\mbox{}\vspace{-\baselineskip}

We briefly recall some properties of Krein spaces. A detailed review is given in Refs.~\cite{Bognar}~-~\cite{Krein}.

If an inner product space $\kr$ admits a fundamental decomposition of the form
\begin{equation}\label{dec}
\kr=\kr^++\kr^-;\quad \kr^+\perp\kr^-,
\end{equation}
where $\kr^+$ and $\kr^-$ are complete subspaces with positive and negative metrics respectively, then we shall say that $\kr$ is a Krein space.

%Loosely speaking, a Krein space is a non-degenerate, decomposable, complete inner product space.

According to the definition, the class of Krein spaces includes Hilbert spaces ($\kr^-=0$) as well as anti-spaces of Hilbert spaces ($\kr^+=0$).

Since every vector in a Krein space admits decomposition
\begin{equation}
x=x_++x_-,\,x_{\pm}\in\kr_{\pm},
\end{equation} the inner product of
$x$ and $y$ can be written as follows
\begin{equation}
(x,y)=(x_+,x_+)+(y_-,y_-).
\end{equation}

We say that $J$ is the fundamental symmetry operator belonging to the fundamental decomposition~\refn{dec} if
\begin{equation}\label{J}
J(x_++x_-)=x_+-x_-.
\end{equation}
The following properties can easily be verified:
\begin{enumerate}
\item $J$ is completely invertible and $J^{-1}=J$.
\item $J$ is  self-adjoint.
\end{enumerate}

Since $J$ is   self-adjoint, the formula
\begin{equation}
(x,y)_J=(x,Jy)=(x_+,y_+)-(x_-,y_-)
\end{equation}
defines a positive scalar product called the $J$-inner product on a Krein space. Indeed,
$$
(x,x)_J=(x_+,x_+)-(x_-,x_-)\geqslant0.
$$

A norm on $\kr$ can be defined by
\begin{equation}
\|x\|_J=+\sqrt{(x,x)_J},\quad x\in\kr
\end{equation}
and is called the $J$-norm.

It is easy to obtain the relation between the operators $A^*$ and $A^+$, where the symbols ``\,*\,'' and ``\,+\,'' denote adjoint operators relating to the $J$-inner product and the indefinite inner product respectively. Indeed, since
\begin{equation}
(Ax,y)=(Ax,Jy)_J=(x,A^*Jy)_J=(x,JA^*Jy),
\end{equation}
it follows that
\begin{equation}\label{krel}
A^+=JA^*J.
\end{equation}

%It is important to note that for every fundamental symmetry $J$ the $J$-inner product turns $\kr$ into a Hilbert space~(\citer{Bognar}).

\section{Regular Heisenberg algebra representations in Krein space}
\mbox{}\vspace{-\baselineskip}

 Suppose that the number operator $N=a^*a$ has an eigenvector $\psi_{\alpha}$  and
\begin{equation}\label{eigen}
N\psi_{\alpha}=\alpha\psi_{\alpha}.
\end{equation}
%For regularity of the CCR representation in Hilbert space it is enough to
Let us prove first that $\alpha\geqslant0$ in Hilbert space. Indeed, if $\alpha=-\beta,\,\beta>0$, then, on the one hand,
\begin{equation}
(N\psi_{-\beta},\psi_{-\beta})=-\beta(\psi_{-\beta},\psi_{-\beta})\leqslant0,
\end{equation}
and, on the other hand,
\begin{equation}
(N\psi_{-\beta},\psi_{-\beta})=(a\psi_{-\beta},a\psi_{-\beta})\geqslant0.
\end{equation}
From these inequalities it follows that $\psi_{-\beta}=0$. Since $a\psi_{\alpha}\sim\psi_{\alpha-1}$, the requirement for $\alpha$ to be non-negative imposes condition~\refn{vac}, i.~e.,
the representation under consideration is a Fock one. Condition~\refn{eigen} can be used as another definition of regularity of representations in Hilbert space.

By definition condition~\refn{eigen} introduces regular representations in spaces with an indefinite metric as well. The following theorem classifies regular representations in Krein spaces.\citev{jmath}
\renewcommand{\thesection}{\arabic{section}}
\begin{theorem}
The regular irreducible representations of the Heisenberg algebra in a Krein space fall into the following classes:
\begin{enumerate}
\item Fock case: it is characterized by the existence of a vector $\psi_0$~(Fock vector) satisfying
$$
a\psi_0=0.
$$
The number operator N has a complete set of eigenvectors $\psi_n,\,n=0,1,2,\ldots$, with  non-negative integer eigenvalues;
\item Anti-Fock case: it is characterized by the existence of a vector $\psi_{-1}$~(anti-Fock vector) satisfying
$$
a^+\psi_{-1}=0.
$$
The number operator N has a complete set of eigenvectors $\psi_{n},\,n=-1,-2,\ldots$, with negative integer eigenvalues;
\item $\Lambda$-case: the number operator N has a complete set of eigenvectors such that
$$
\mathrm{Sp}N=\lambda+\mathbb Z,\quad-1<\lambda<0.
$$
\end{enumerate}
\end{theorem}
The Fock representations in Krein space are equivalent to those in Hilbert space. The Weyl representation for the CCR algebra was proved to exist in this case.\citev{Foi,Vernreg} Now we are interested in the anti-Fock case that implies
\begin{equation}\label{kgr}
a^+\psi_{-1}=0.
\end{equation}
It is easy to see that
\begin{equation}\label{norm}
(\psi_{-n},\psi_{-n})=(-1)^{n-1}\,\,,
\end{equation}
where the set of normalized eigenvectors $\{\psi_{-n}\}$ is defined by
$$
\psi_{-n}=\frac{a^{n-1}}{\sqrt{(n-1)!}}\,\psi_{-1},\quad n\in\mathbb N.
$$
Indeed,
\begin{equation*}
\begin{split}
(\psi_{-n},\psi_{-n})&=\frac1{n-1}(a\psi_{-n+1},a\psi_{-n+1})\\
&=\frac1{n-1}(\psi_{-n+1},a^+a\psi_{-n+1})=(-1)(\psi_{-n+1},\psi_{-n+1})\\
&=\ldots=(-1)^{n-1}(\psi_{-1},\psi_{-1}),
\end{split}
\end{equation*}
where we can always take $(\psi_{-1},\psi_{-1})=1$.

Let us prove that
\begin{equation}\label{antic}
\{a,J\}=\{a^+,J\}=0,
\end{equation}
where $\{x,y\}=xy+yx$, is satisfied on linear combinations of vectors $\psi_{-n}$.
Let $n=2m$. Since $a\psi_{-2m}=\psi_{-2m-1}$, in view of~\refn{J} and~\refn{norm} it is clear that
$$
Ja\psi_{-2m}=J\psi_{-2m-1}=\psi_{-2m-1}.
$$
On the other hand,
$$
aJ\psi_{-2m}=-a\psi_{-2m}=-\psi_{-2m-1}.
$$
The case $n=2m+1$ can be considered similarly. Thus we have shown that
$$
\{a,J\}\psi_{-n}=0,\quad n\in\mathbb N.
$$
Next, for every vector $\psi$ such that
$$
\psi=\sum_{p=-k}^{-l} c_p\,\psi_p,\quad k,l\in\mathbb N,\quad k>l\geqslant1,
$$
it follows that
\begin{equation}\label{den}
\{a,J\}\psi=0.
\end{equation}
Thus $\{a,J\}=0$ is satisfied on a dense domain in a Krein space. Since $J^+=J$, the relation $\{a^+,J\}=0$ is satisfied on the domain as well.

\renewcommand\thesection{\Roman{section}.\!\!\!\!\!}
%\newpage
\section{Weyl representation analogue }
\mbox{}\vspace{-\baselineskip}

In this section we show how an analogue of the Weyl representation can be derived for the case of an anti-Fock realization on a Krein space.

Let $a$ and $a^+$ be the anti-Fock representation operators on a Krein space $\kr$. Let us introduce the operators $b=a^+$ and $b^+=a$  and obtain
\begin{equation}
[b,b^+]=-1,\quad \tilde{N}=-N-1,\quad \mathrm{Sp}\tilde{N}=\mathbb N,\quad\psi_{-n}=\tps_{n-1},
\end{equation}
where ``\,\~ \,'' denotes the corresponding notions defined in terms of the new operators.

Relations~\refn{kgr} and~\refn{antic} take form
\begin{equation}
b\tps_0=0,\quad \{b,J\}=\{b^+,J\}=0,
\end{equation}
hence, by~\refn{krel},
\begin{equation}
b^*=Jb^+J=-b^+.
\end{equation}
Note that the operators $b$ and $b^*$ satisfy the standard commutation relation of the form~\refn{aa}
\begin{equation}
[b,b^*]=I.
\end{equation}

In view of~\refn{aa*} the operators $\tp$ and $\tq$ satisfying the canonical commutation relation~\refn{CCR} can be expressed by
\begin{equation}\label{f}
\tp=\frac1{\sqrt2i}(b-b^*),\quad\tq=\frac1{\sqrt2}(b+b^*).
\end{equation}
It is easy to express them in terms of the original anti-Fock operators as well. Indeed, making use of the relation $b^*=-b^+=-a,\,b=a^+$, we obtain that
\begin{equation}
\tp=\frac1{\sqrt2i}(a^++a),\quad\tq=\frac1{\sqrt2}(a^+-a).
\end{equation}
We see that $\tp$ and $\tq$ are self-adjoint operators defined on some domain in a Hilbert space. We can repeat the arguments used in~\citer{Put} in the proof of eq. (\ref{Weyl}). Thus we obtain an analogue of the Weyl form of the CCR representation for the anti-Fock case.

Finally, we can apply von Neumann's theorem to this representation. Indeed, unitary groups $\{U(t)=e^{it\tp}\}$ and $\{V(s)=e^{is\tq}\}$ of operators satisfy all the conditions of the theorem.

\renewcommand\thesection{\Roman{section}.\!\!\!\!\!}

\section{Conclusions}
\mbox{}\vspace{-\baselineskip}

So far we have considered quantum systems with only a finite number $N$ of degrees of freedom, where von Neumann's theorem guarantees the unitary equivalence of any two irreducible canonical systems with equal $N$. Contrariwise, in the case of field theories the Haag's theorem\citev{SW,BLT} shows that  free and interacting quantum fields cannot  be related by a unitary operator.  We hope that the results presented in this paper can be a preparatory step to the consideration of the CCR representations in Krein space, when the number of degrees of freedom is infinite, and to the generalization of Haag's theorem to the case of non-physical particles.

\end{document}